\documentclass{aa}
\usepackage[varg]{txfonts}
\usepackage{graphicx}
\usepackage{natbib}
\bibpunct{(}{)}{;}{a}{}{,} 
\usepackage{color}
\usepackage{amstext}

\begin{document}

\title{Solar activity over nine millennia: A consistent multi-proxy reconstruction}

\author{Chi Ju Wu\inst{1}
\and I. G. Usoskin\inst{2,3}
\and N. Krivova\inst{1}
\and G. A. Kovaltsov\inst{4}
\and M. Baroni\inst{5}
\and E. Bard\inst{5}
\and S. K. Solanki\inst{1,6}
}

\institute{Max-Planck-Institut f\"ur Sonnensystemforschung, G\"ottingen, Germany
\and Space Climate Research Unit, University of Oulu, Finland
\and Sodankyl\"a Geophysical Observatory, University of Oulu, Finland
\and Ioffe Physical-Technical Institute, 194021 St. Petersburg, Russia
\and CEREGE, Aix-Marseille University, CNRS, Coll\`ege de France, Technop\^ole de l'Arbois, Aix-en-Provence, France
\and School of Space Research, Kyung Hee University, Yongin, Gyeonggi-Do,446-701, Republic of Korea
}

\date{}

\abstract {}
{The solar activity in the past millennia can only be reconstructed from cosmogenic radionuclide proxy records in terrestrial archives.
However, because of the diversity of the proxy archives, it is difficult to build a homogeneous reconstruction.
All previous studies were based on individual, sometimes statistically averaged, proxy datasets.
Here we aim to provide a new consistent multi-proxy reconstruction of the solar activity over the last 9000 years,
 using all available long-span datasets of $^{10}$Be and $^{14}$C in terrestrial archives.
}
{A new method, based on a Bayesian approach, was applied for the first time to solar activity reconstruction.
A Monte Carlo search (using the $\chi^2$ statistic) for the most probable value of the modulation
 potential was performed to match data from different datasets for a given time.
This provides a straightforward estimate of the related uncertainties.
We used six $^{10}$Be series of different lengths (from 500--10000 years) from Greenland and Antarctica, and the global $^{14}$C production
 series.
The $^{10}$Be series were resampled to match wiggles related to the grand minima in the $^{14}$C reference dataset.
The stability of the long data series was tested.
}
{The Greenland Ice-core Project (GRIP) and the Antarctic EDML
 (EPICA Dronning Maud Land) $^{10}$Be series diverge
 from each other during the second half of the Holocene, while the $^{14}$C series lies in between them.
A likely reason for the discrepancy is the insufficiently precise beryllium
 transport and deposition model for Greenland, which leads to an undercorrection of the GRIP series for
 the geomagnetic shielding effect.
A slow 6--7-millennia variability with lows at ca. 5500 BC and 1500 AD in the long-term evolution
 of solar activity is found.
Two components of solar activity can be statistically distinguished: the main component, corresponding
 to the `normal' moderate level, and a component corresponding to grand minima.
A possible existence of a component representing grand maxima is indicated, but
 it cannot be separated from the main component in a statistically significant manner.
 }
{A new consistent reconstruction of solar activity over the last nine millennia is presented
 with the most probable values of decadal sunspot numbers and their realistic uncertainties.
Independent components of solar activity corresponding to the main moderate activity and the grand-minimum state
 are identified; they may be related to different operation modes of the dynamo.
}

\maketitle

\section{Introduction}
The Sun is an active star whose magnetic activity varies on different timescales, from seconds to millennia.
Understanding solar variability in detail is important for many reasons, ranging from applications in stellar astrophysics
 and dynamo theory to paleoclimatic and space weather studies.
Various direct, partly multiwavelength spectroscopic solar observations cover the past few decades up to a
 century in the past.
They provide knowledge of the solar variability that is expressed in different indices.
For the preceding centuries back to 1610, only visual
 information from simple optical observations in the form of sunspot numbers (SNs) is available \citep{hathawayLR}.
The quality of the SN series varies, and the uncertainty increases sufficiently before the nineteenth
 century \citep{clette14,usoskin_LR_17}.
However, the Maunder minimum in the second half of the seventeenth century was observed sufficiently well to conclude that
 sunspot activity was exceptionally low \citep{ribes93,vaquero15,usoskin_MM_15}.
For earlier times, only indirect proxies can help assessing solar activity
 \footnote{Although naked-eye observations of sunspots are available, they do not provide quantitative assessments
  \citep{usoskin_LR_17}.}.
Such proxies are, for example, concentrations of cosmogenic radionuclides radiocarbon ($^{14}$C),
 beryllium-10 ($^{10}$Be), or chlorine-36 ($^{36}$Cl) that are
measured in tree trunks or polar ice cores,
 respectively.
These archives are dated independently.
The use of cosmogenic proxies for studying the solar activity in the past has been proposed long ago
 \citep[e.g.,][]{stuiver61,stuiver80,beer88}, and the method has been developed since then in both
 measurements and modeling \citep[see reviews by][and references therein]{beer12,usoskin_LR_17}.
These data cover timescales of up to ten millennia and more.

Cosmogenic radionuclides are produced as a byproduct of a nucleonic cascade initiated by galactic cosmic rays (GCR)
 in the Earth's atmosphere.
This is the only source of these nuclides in the terrestrial system (which is why they are called ``cosmogenic'').
Radiocarbon $^{14}$C is mainly produced as a result of neutron capture ($np-$reaction) by nitrogen,
 which is responsible for $>99$\% of the natural $^{14}$C production.
The isotope $^{10}$Be is produced as a result of spallation reactions of O and N nuclei caused by energetic GCR particles.
After production, the two radionuclides have different processes of transport, deposition, and storage in terrestrial
 archives around the globe.
Radiocarbon is mostly measured in dendrochronologically dated annual rings of live or dead tree trunks, while $^{10}$Be
 is measured in glaciologically dated polar ice cores mostly from Greenland or Antarctica.
In addition to the geomagnetic field, the flux of GCRs impinging on Earth is modulated by large-scale heliospheric
 magnetic features (interplanetary magnetic fields and solar wind) so that the measured content of the nuclides may serve as a
 proxy of solar magnetic variability in the past (after the influence of the geomagnetic field has been removed).
Different cosmogenic isotope series exhibit a high degree of similarity on timescales of a century to a millennium because
they share the same production origin \citep{bard97, vonmoos06, beer12, usoskin_10Be_09, delaygue11, usoskin_AA_16, adolphi16}.
However, systematic discrepancies between long-term (multi-millennial) trends in different series can be observed
 \citep{vonmoos06, inceoglu15, adolphi16,usoskin_AA_16}, probably because of the influence of climate conditions
  (regional deposition pattern
 for $^{10}$Be or large-scale ocean circulation for $^{14}$C) or an improper account for the geomagnetic field variation,
to which the two isotopes respond differently.

Because of the differences in the isotope records, earlier reconstructions of solar activity were obtained based
 on individual cosmogenic series, leading to a diversity in the results.
Earlier multi-proxy efforts
 \citep[e.g.,][]{bard97,mccracken04,vonmoos06,muscheler07,usoskin_AA_07,knudsen_GRL_09} were mostly based on
 a simple comparison of individual records.
The first consistent effort to produce a merged reconstruction was made by \citet{steinhilber12}, who used
 the principle component analysis (PCA) to extract the common variability signal (assumed to be solar) from the reconstructions
 based on three cosmogenic series (global $^{14}$C, Antarctic EDML (EPICA Dronning Maud Land), and the
 Greenland Ice-core Project, GRIP, $^{10}$Be) and to
 remove the system effects (e.g., the deposition process, snow accumulation rate, and changes in the carbon cycle
 and dating uncertainties), which are different for each series.
The PCA method keeps only the relative variability and looses the information on the absolute level, which needs further normalization.
Moreover, this method effectively averages multiple signals without taking the accuracy of each data point and
 possible time lags between the signals into account \citep[see Figures S1--S8 in][]{steinhilber12}.
As discussed by \citet{adolphi16}, the time mismatch between
the $^{14}$C and $^{10}$Be series may be as large as
 70 years toward the early Holocene, however.

We here introduce a new method for a consistent multi-proxy reconstruction of the solar activity that is based on the
Bayesian approach to determine the most probable value (and its uncertainties) of the solar activity at any moment
 in time by minimizing the $\chi^2$-discrepancy between the modeled and the actually measured
 cosmogenic isotope data.
This method straightforwardly accounts for error propagation and provides the most probable reconstruction and its
 realistic uncertainties.
Since the method is sensitive to the dating accuracy of different records, we redated the $^{10}$Be records to match
 their dating with that of $^{14}$C by applying the standard wiggle-matching method \citep{cain76,muscheler14}
to the official ice-core chronology, as described in Section~\ref{sec:prep}.
The solar modulation potential was assessed using the Bayesian approach from all the
 available datasets (Section~\ref{sec:phi_recon}).
Finally, the SN series was calculated (Section~\ref{sec:SN}).
Section 6 summarizes our results and conclusions.

\section{Data}   
\subsection{Cosmogenic isotope records}
\begin{figure}
\centering
\includegraphics[width=1.0\columnwidth]{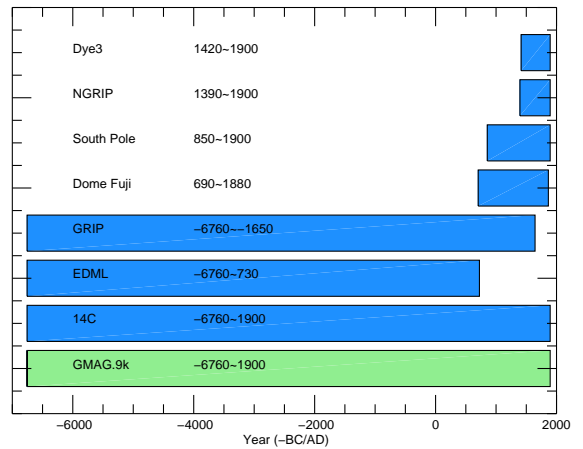}
\caption{Data series (see Table~\ref{tbl:series}) and temporal coverage.
Blue shading denotes the cosmogenic isotope series, while green shading shows the geomagnetic series.}
\label{fig:series}
\end{figure}
We here used six $^{10}$Be series from Greenland and Antarctica and one global $^{14}$C production
 series, as summarized in Table\ref{tbl:series} and Figure~\ref{fig:series}.

The $^{14}$C series covers the entire Holocene with homogeneous resolution as given by the globally averaged
 $^{14}$C production rate computed by \citet{roth13} from the original International $^{14}$C Calibration dataset
 INTCAL09 \citep{reimer09} measurements of $\Delta^{14}$C.
While this series has a pseudo-annual temporal sampling \citep{roth13}, its true resolution is decadal.
The series is provided as an ensemble of 1000 realizations of individual reconstructions.
Each realization presents one possible reconstruction in the sense of a Monte Carlo approach (viz. one realized path
 of all possible paths in the parametric space), considering all known uncertainties.
This enables directly assessing the error propagation throughout the entire process.

$^{10}$Be series have different coverage and temporal resolutions.
We reduced them to two cadences.
Annual series were kept as they are, while rougher resolved series were resampled to decadal cadence.
For each series, we also considered 1000 realizations that were
synthesized using the mean curve and the standard deviation (error bars).
Two series were updated with respect to earlier studies.
The EDML series was used with the new Antarctic Ice Core Chronology,
version 2012
(AICC2012) \citep{veres13,bazin13}.
The GRIP series was updated from its original ss09(sea) timescale \citep{johnsen95,johnsen01} to the more recent
 timescale of the Greenland Ice Core Chronology, version 2005 (GICC05) \citep{vinther06}.

All $^{10}$Be series were converted into units of production/flux [atoms cm$^{-2}$ sec$^{-1}$] when possible,
which is
 natural for the isotope production by cosmic rays.
Originally, $^{10}$Be measurements are given in units of concentration [atoms g$^{-1}$].
To convert them into the flux, the independently measured or
obtained snow accumulation rate at each site is needed.
The accumulation rate was considered individually for each ice core, using the same chronology
 as for the $^{10}$Be data (\citet{veres13,bazin13} for EDML on AICC2012 and \citet{rasmussen14,seierstad14}
 for GRIP on the GICC05 timescale).
For the Dye-3 and South Pole series, accumulation data are not provided, and in these cases, we used concentration
 data assuming that the concentration is proportional to the depositional flux, with the scaling factor being a
 free parameter (Table~\ref{tbl:series}).

\begin{table*}
\caption{ Temporal coverage, cadence, type of data (production rate, PR; depositional flux, D; or concentration, C)
 and the resulting best-fit scaling factors, $\kappa$ (Section~\ref{sec:phi_recon}), of the cosmogenic isotope series.}
\label{tbl:series}
\begin{tabular}{l | c c l c c c}
\hline
Series & Location & Period (-BC/AD) & Cadence & Data & $\kappa$ & Reference$^\ddagger$ \\
\hline
$^{14}$C: INTCAL09 & Global & -8000 -- 1950 & Decadal & PR & -- & RJ2013 \\
$^{10}$Be: GRIP & Greenland & -7375 -- 1645 & Decadal $^\dagger$ & D & 1.028 & Yea1997, Mea2004, Vea2006 \\
$^{10}$Be: EDML & Antarctica & -7440 -- 730 & Decadal $^\dagger$ & D & 0.815 & Sea2012 \\
$^{10}$Be: NGRIP & Greenland & 1389 -- 1994 & Annual & D & 0.87 & Bea2009 \\
$^{10}$Be: Dye3 & Greenland & 1424 -- 1985 & Annual & C & 94.4 & Bea1990, Mcea2004 \\
$^{10}$Be: Dome Fuji (DF) & Antarctica & 690 -- 1880 & 5-year $^\dagger$ & D & 0.87 & Hea2008 \\
$^{10}$Be: South Pole (SP) & Antarctica & 850 -- 1960 & Decadal & C & 342.2 & Rea1990, Bea1997\\
\hline
\end{tabular}
\\$^\dagger$ Resampled to decadal resolution.
\\$^\ddagger$ References: RJ2013 \citep{roth13}; Yea1997 \citep{yiou97}; Mea2005 \citep{muscheler04};
 Mcea2004 \citep{mccracken04}; Vea2006 \citep{vonmoos06}; Sea2012 \citep{steinhilber12};
 Bea2009 \citep{berggren09}; Bea1990 \citep{beer90}; Hea2008 \citep{horiuchi08}; Rea1990 \citep{raisbeck90};
 Bea1997 \citep{bard97}.
\end{table*}

\subsection{Geomagnetic data}

As the geomagnetic data for the last millennia, we used the recent archeo/paleomagnetic model GMAG.9k of the
 virtual axial dipole moment (VADM) as published by \citet{usoskin_AA_16} for the period since ca. 7000 BC.
This model provides an ensemble of 1000 individual VADM reconstructions that include all the uncertainties.
The range of the ensemble reconstruction covers other archeo-
and paleomagnetic models
 \citep[e.g.,][]{genevey08,knudsen08,licht13,nilsson14,pavon14}, as is shown in Figure~2 of
 \citet{usoskin_AA_16}, which means that the related uncertainties
are covered.
The length of the geomagnetic series limits our study to the period since 6760 BC.

\section{Data processing}
\label{sec:prep}
\subsection{Temporal synchronization of the records: wiggle matching}
\label{sec:wiggle_matching}

While the radiocarbon is absolutely dated via dendrochronology by tree-ring counting, dating of ice cores is less precise.
In principle, dating of ice cores can be done, especially on short timescales, by counting annual layers of sulfate or sodium when the accumulation and the characteristics of the site allow it \citep[e.g.,][]{sigl15}.
In practice, however, dating of long series is performed by applying ice-flow models between tie points that are related to known
 events, such as volcano eruptions that leave clear markers in ice.
While the accuracy of dating is quite good around the tie points, the exact ages of the samples between the tie points may be rather
 uncertain. This ranges from several years during the last millennium to up to 70--100 years in the earlier part of the Holocene
 \citep{muscheler14,sigl15,adolphi16}.

Since our method is based on the minimization of the $\chi^2$-statistics of all series for a given moment in time,
 it is crucial that these series are well synchronized.
Accordingly, we performed a formal synchronization of the series based on the wiggle-matching procedure,
 which is a standard method for synchronizing time series \citep[e.g.,][]{cain76,hoek01,muscheler14}.
We took the $^{14}$C chronology as the reference and adjusted the timing of all other series to it.

\subsubsection{Choice of wiggles}
\label{Sec:wiggles}
\begin{table*}
\centering
\caption{Wiggles (central date and the length in years) used for synchronization of the various $^{10}$Be time series to $^{14}$C
 \citep{eddy76,stuiver80,stuiver89,goslar03,inceoglu15,usoskin_AA_16}, and the synchronization time
 (in years) for the GRIP and EDML series.
All dates are given in -BC/AD for dendrochronologically dated $^{14}$C.}
\begin{tabular}{cc|cc|cc|cc}
\hline
Date & Length & $dT$ (EDML) & $dT$ (GRIP) & Date & Length & $dT$ (EDML) & $dT$ (GRIP) \\
\hline
1680 & 80 & N/A & N/A & -3325 & 90 & $-20_{-11}^{+10}$ & $7_{-7}^{+6}$ \\
1480 & 160 & N/A & $8_{-6}^{+5}$ & -3495 & 50 & $-23_{-7}^{+6}$ & $6_{-6}^{+6}$ \\
1310 & 80 & N/A & $2_{-5}^{+4}$ & -3620 & 50 & $-26_{-10}^{+21}$ & $5_{-6}^{+6}$ \\
1030 & 80 & N/A & $0_{-4}^{+4}$ & -4030 & 100 & $-19_{-6}^{+6}$ & $2_{-4}^{+5}$ \\
900 & 80 & N/A & $13_{-3}^{+4}$ & -4160 & 50 & $-22_{-6}^{+5}$ & $-3_{-5}^{+5}$ \\
690 & 100 & $16_{-6}^{+8}$ & $-1_{-5}^{+5}$ & -4220 & 30 & $-24_{-5}^{+5}$ & $-6_{-5}^{+5}$ \\
260 & 80 & $12_{-6}^{+7}$ & $0_{-4}^{+5}$ & -4315 & 50 & $-26_{-5}^{+4}$ & $4_{-3}^{+3}$ \\
-360 & 120 & $6_{-6}^{+5}$ & $2_{-3}^{+6}$ & -5195 & 50 & $-19_{-5}^{+4}$ & $7_{-5}^{+5}$ \\
-750 & 70 & $-18_{-9}^{+12}$ & $0_{-5}^{+2}$ & -5300 & 50 & $-18_{-5}^{+4}$ & $7_{-5}^{+5}$ \\
-1385 & 80 & $-25_{-7}^{+9}$ & $3_{-5}^{+4}$ & -5460 & 40 & $-22_{-5}^{+5}$ & $18_{-4}^{+5}$ \\
-1880 & 80 & $-30_{-6}^{+7}$ & $5_{-6}^{+6}$ & -5610 & 40 & $-36_{-4}^{+9}$ & $29_{-8}^{+7}$ \\
-2120 & 40 & $-27_{-6}^{+5}$ & $6_{-6}^{+7}$ & -5970 & 60 & $-29_{-8}^{+14}$ & $20_{-6}^{+6}$ \\
-2450 & 40 & $-28_{-5}^{+4}$ & $0_{-3}^{+4}$ & -6060 & 60 & $-27_{-9}^{+13}$ & $15_{-8}^{+6}$ \\
-2570 & 100 & $-29_{-4}^{+4}$ & $0_{-3}^{+3}$ & -6385 & 130 & $-22_{-11}^{+10}$ & $-4_{-14}^{+6}$ \\
-2855 & 90 & $-20_{-6}^{+7}$ & $-14_{-10}^{+16}$ & -6850 & 100 & $-19_{-7}^{+7}$ & $-24_{-7}^{+8}$ \\
-3020 & 60 & $-18_{-8}^{+6}$ & $-7_{-9}^{+12}$ & -7030 & 100 & $-25_{-6}^{+6}$ & $-39_{-12}^{+8}$ \\
-3080 & 60 & $-32_{-6}^{+6}$ & $-4_{-9}^{+11}$ & -7150 & 100 & $-29_{-5}^{+5}$ & $-49_{-15}^{+8}$ \\
\hline
\end{tabular}
\label{Tab:wiggles}
\end{table*}

\begin{figure}[t]
\centering
\includegraphics[width=1.0\columnwidth]{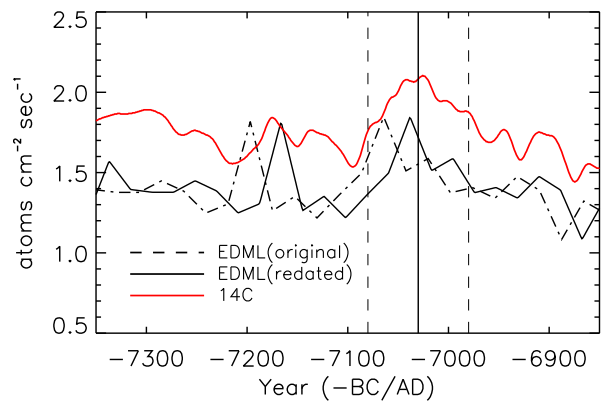}
\caption{Typical example of a wiggle in the $^{14}$C (red) series compared with the original EDML
 $^{10}$Be series (dashed black, scaled up by a factor of 172).
Solid and dashed vertical lines denote the middle and the span of the wiggle considered for $^{14}$C.
The black sold curve shows the EDML $^{10}$Be series after the synchronization (see Sect. \ref{sec:atfs}).}
\label{fig:pickexamples}
\end{figure}

Since cosmogenic isotope series exhibit strong fluctuations of various possible origins, it is important to
 match only those wiggles that can presumably be assigned to the production changes, that is, cosmic ray (solar)
 variability, and exclude possible regional climate spells.
One well-suited type of wiggles is related to the so-called grand minima of solar activity, which are
 characterized by a fast and pronounced drop in solar activity for several decades.
The most famous example of a grand minimum is the Maunder minimum,
which occurred between 1645--1715 \citep{eddy76,usoskin_MM_15}.
Grand minima can be clearly identified in the cosmogenic isotope records as sharp spikes \citep{usoskin_AA_07,inceoglu15}.
A typical (although not the most pronounced) example of such wiggles (grand minima) is shown in Figure~\ref{fig:pickexamples}
 for the (scaled) original EDML $^{10}$Be series in a dashed curve along with the $^{14}$C production data in red.
Although the overall variability of the two series looks similar, the mismatch in the timing
 between them is clear and is roughly a few decades.

For the further analysis, we selected periods with clear wiggles (spikes, corresponding to grand minima, {as listed
 in earlier works -- see references in the notes to Table~\ref{Tab:wiggles}}) in the $^{14}$C
 series as listed in Table~\ref{Tab:wiggles}, along with their centers and time span.

\subsubsection{Synchronization of the wiggles}
\label{sec:atfs}
\begin{figure}[t]
\centering
\includegraphics[width=1\columnwidth]{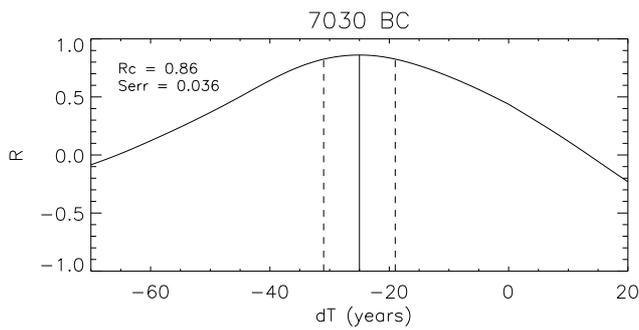}
\caption{Example of the calculation of the best-fit time adjustment $dT=-25$ years (solid vertical line)
 and its 68\% confidence interval (dashed lines, -31 and -19 years) for the wiggle case shown in Figure~\ref{fig:pickexamples}.
 The Pearson linear correlation was calculated between the $^{14}$C and EDML $^{10}$Be series for the $\pm 100$-year time
  window around the center of the wiggle.}
\label{fig:uncertaintysample}
\end{figure}

For all the selected wiggles (Table~\ref{Tab:wiggles}), we found the best-fit time adjustment $dT$
 between the analyzed $^{10}$Be and the reference $^{14}$C production series by maximizing the cross correlation
 between the series, calculated within a time window centered at the middle of the wiggle.
The data were annually interpolated within the time windows so that the time step in defining $dT$ was one year.
The length of the correlation window was chosen as twice the length of the wiggle (see Table~\ref{Tab:wiggles}).
For each wiggle, we repeatedly calculated the Pearson linear correlation coefficients between the $^{14}$C and $^{10}$Be series, and we selected the value of $dT$  that maximized the cross-correlation coefficient $R$ between the two series.
The standard error ($s_{\rm err}$) of the correlation coefficient was calculated using the approximate formula
\citep[e.g.,][]{cohen03}:
\begin{equation}
s_{\rm err} = \sqrt{ \frac{1-R_c^2}{n-2}},
\end{equation}
where $R_c$ is the maximum correlation coefficient and $n$ is the number of the data points within the correlation window.
This uncertainty $s_{\rm err}$ was translated into the $1\sigma$ confidence interval for $dT$, as
 illustrated in Figure~\ref{fig:uncertaintysample}, which shows the correlation coefficient, $R$ (black curve),
  as a function of the time shift $dT$.
It reaches its maximum $R_c=0.86$ at $dT=-25$ years, as indicated by the vertical solid line.
The dotted lines bound the 68\% confidence interval for $dT$ defined as $R=R_c-s_{\rm err}$.

The `momentary' time adjustments $dT$ were considered as `tie points' (listed in Table~\ref{Tab:wiggles})
 for the $^{10}$Be series, with a linear interpolation used between them.
Adjustments for the EDML series, based here on a new Antarctic Ice Core Chronology \citep[AICC2012;][]{veres13},
 lie within +20/-40 years, while for the GRIP series, they vary within +20/-50 years.
The adjustment range for GRIP is concordant with the Greenland chronology correction function
 \citep[e.g.,][]{muscheler14} within the uncertainties of the GICC05 timescale \citep{seierstad14}.
Some discrepancies may be caused by the differences in the datasets and applied method.
There are no earlier results for the EDML synchronization chronology to be compared with.
We emphasize here that we do not pretend to perform a full chronological scale update, but only to match wiggles
 between a single beryllium series and $^{14}$C data, which is sufficient for this work.
In particular, earlier synchronization studies produced smooth correction curves \citep[e.g.,][]{knudsen_GRL_09,muscheler14},
 where individual wiggles may still be slightly mismatched, while we are focused here on matching each
 wiggle in each series separately.

An example of the resulting redated $^{10}$Be series compared to the reference $^{14}$C record is
 shown with the solid black curve in Figure~\ref{fig:pickexamples}.
The synchronization obviously improves the cross correlation between the series, as shown in Table~\ref{Tab:Rc}.
The improvement (in terms of the ratio of $R^2$, which is a measure of the power of covariability between the original
 and synchronized series) is significant for the long
 (1.1 and 2.14 for the GRIP and EDML series, respectively) and shorter Greenland series (NGRIP and Dye3), but small
 for the short Antarctic series.
\begin{table}
\centering
\caption{Squared correlation coefficients between the six $^{10}$Be series and the $^{14}$C series for the
 originally dated ($R_{\rm o}$) and synchronized ($R_{\rm s}$) series.
 The improvement factor $f$ is defined as the ratio of the squared correlation coefficients.}
\begin{tabular}{c c c c c c c}
\hline\hline
   & GRIP & EDML & NGRIP & Dye3 & DF & SP \\
\hline
$R_{\rm o}^2$ & 0.60 & 0.18 & 0.12 & 0.23 & 0.57 & 0.53 \\
$R_{\rm s}^2$ & 0.66 & 0.38 & 0.14 & 0.26 & 0.59 & 0.53 \\
$f$ & 1.11 & 2.14 & 1.11 & 1.14 & 1.03 & 1.01 \\
\hline
\end{tabular}
\label{Tab:Rc}
\end{table}

\begin{figure}[ht!]
\includegraphics[width=0.9\columnwidth]{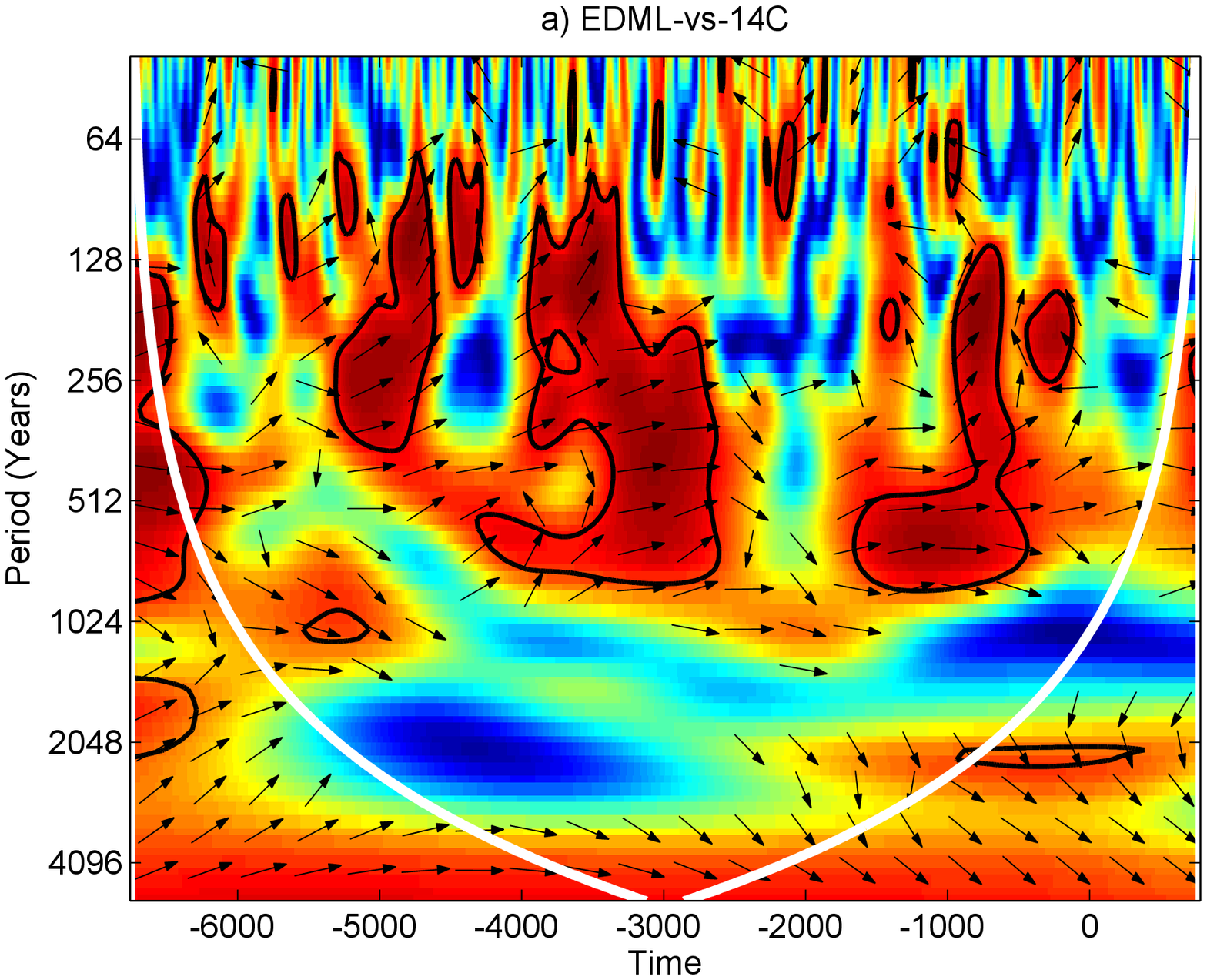}
\vskip 0.2cm
\includegraphics[width=0.9\columnwidth]{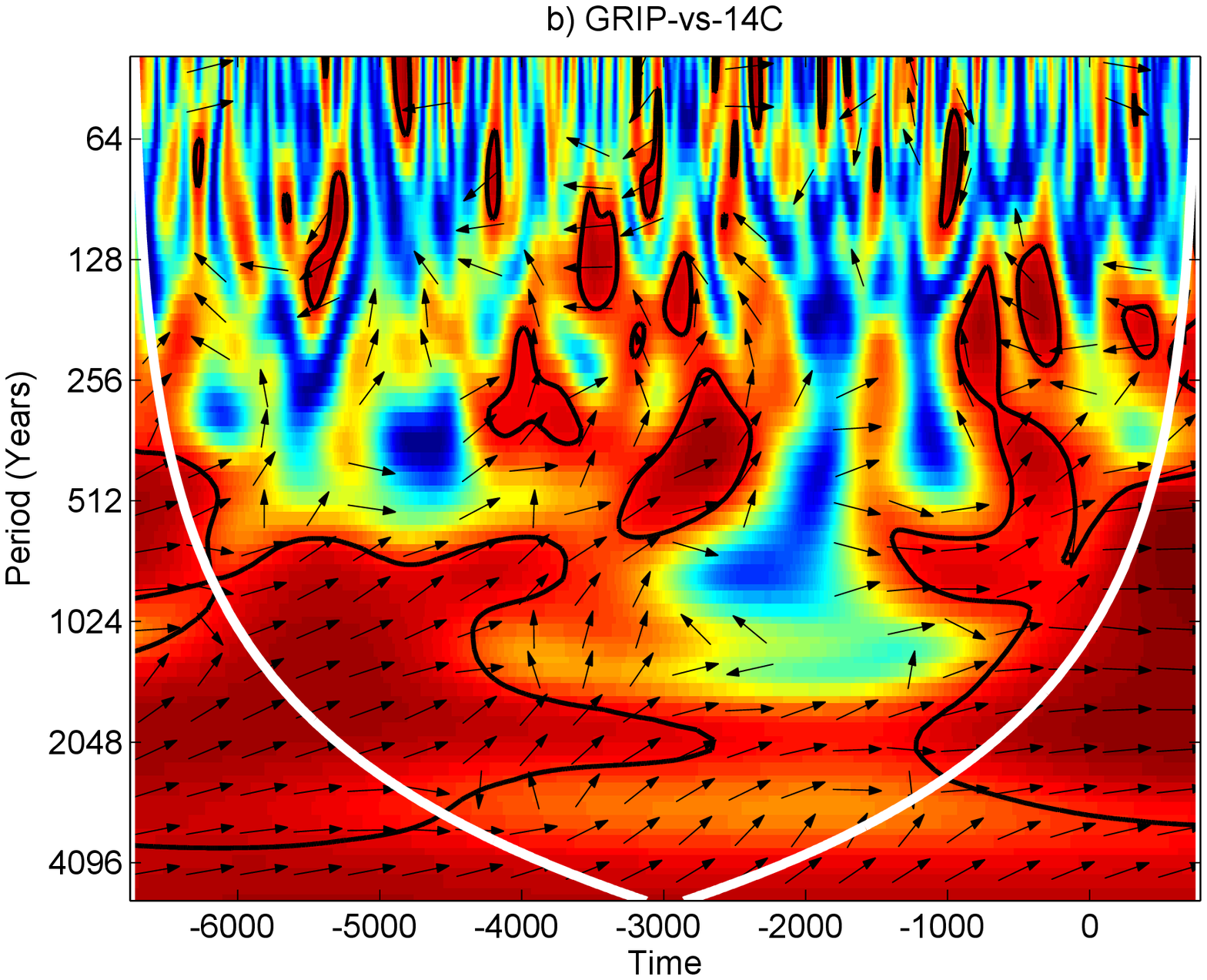}
\vskip 0.2cm
\includegraphics[width=0.9\columnwidth]{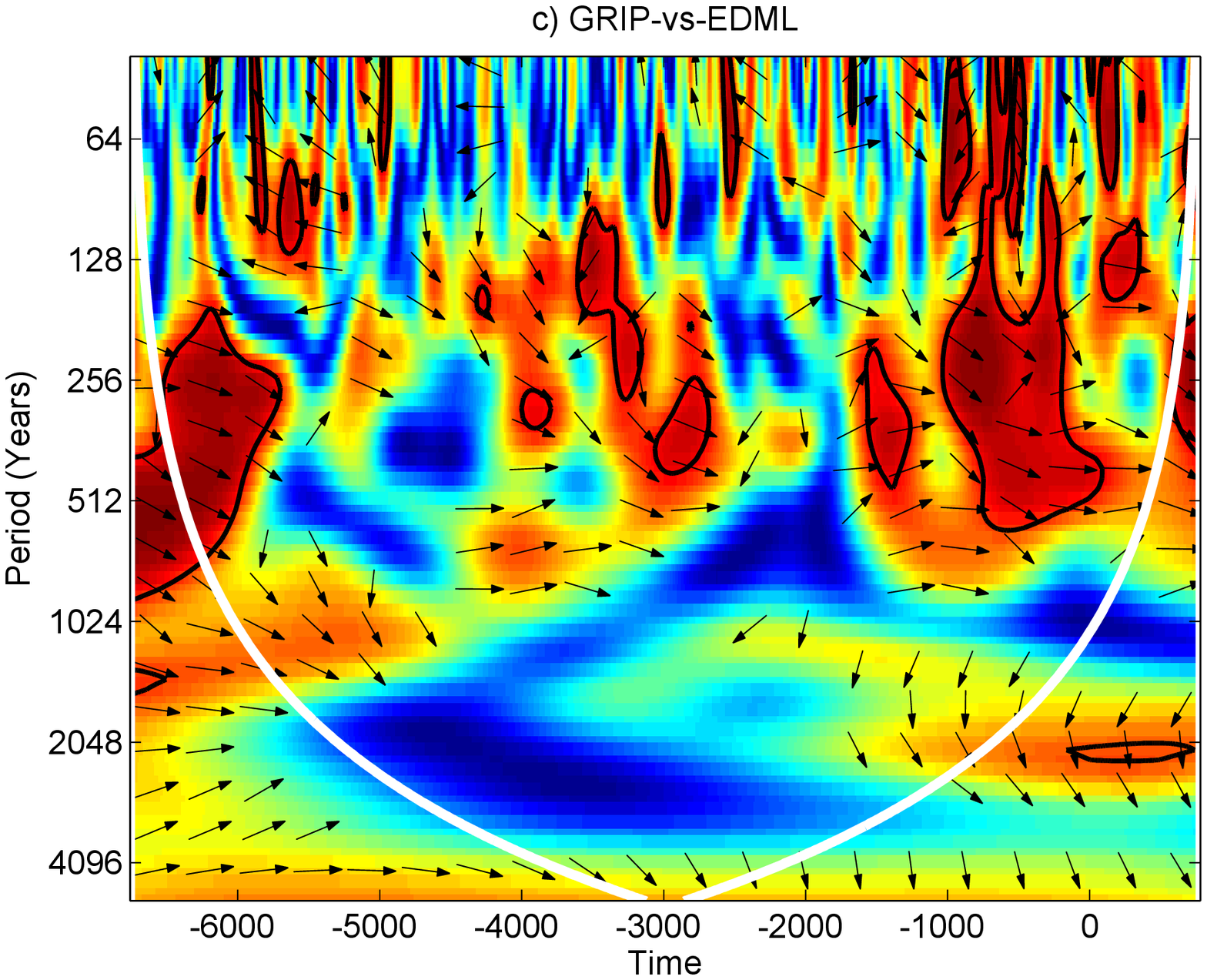}
\includegraphics[width=0.09\columnwidth]{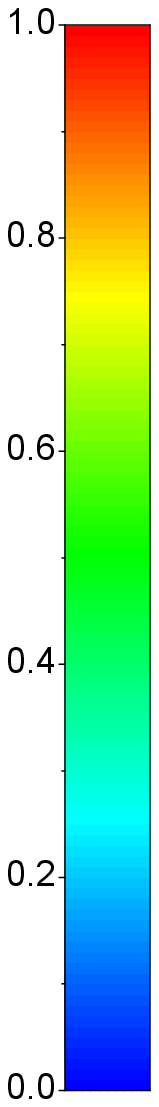}
\caption{Wavelet coherence between long series, redated using the wiggle matching, considered here:
(a) $^{10}$Be EDML vs. $^{14}$C,
(b) $^{10}$Be GRIP vs. $^{14}$C, and
(c) $^{10}$Be GRIP vs. $^{10}$Be EDML.
 The color scale ranges from 0 (deep blue) to 1 (dark red).
 Arrows denote the relative phasing between the series: arrows pointing right denote phase matching,
  while arrows pointing left show the antiphase.
 The white curves denote the cone of influence (COI) beyond which the result is unreliable.}
\label{fig:ww_a}
\end{figure}

The pairwise wavelet coherence between long-term series is shown in Figure~\ref{fig:ww_a},
calculated following the procedure described in \citet{usoskin_10Be_09}, including the
 significance estimate using the non-parametric random-phase method \citep{ebisuzaki97}.
The coherence between EDML $^{10}$Be and $^{14}$C series (panel a) is good
 on timescales shorter than 1000 years and insignificant on timescales longer than 2000--3000 years,
 with no coherence in between.
The coherence between the GRIP $^{10}$Be and $^{14}$C series is good at all timescales longer than 400--500 years.
The coherence between the GRIP and EDML series is intermittent on timescales shorter than 1000 years and
 insignificant on longer timescales.

\section{Reconstruction of the solar modulation potential}
\label{sec:phi_recon}

Since cosmogenic isotopes are produced by cosmic rays in the Earth's atmosphere \citep{beer12}, their measured
 production/depositional flux reflects changes in the cosmic ray flux in the past.
In turn, cosmic rays are modulated by solar magnetic activity, which is often quantified in terms of the modulation
 potential $\phi$.
The latter is a useful parameter to describe the solar modulation of GCRs using the so-called force-field
 parametrization formalism \citep[e.g.,][]{caballero04,usoskin_Phi_05}.
In an ideal case, when both the production rate of cosmogenic isotopes and the geomagnetic field at a given time are known,
 the corresponding modulation parameter can be calculated for the given isotope using a production model that considers
 in great detail all the processes of the nucleonic-muon-electromagnetic cascade that are triggered by energetic cosmic rays in the atmosphere.
Here we used the production model by \citet{poluianov16}, which is a recent update of the widely used cosmic ray atmospheric cascade model
 \citep[CRAC;][]{kovaltsov10,kovaltsov12}.
This model provides absolute production rates and is in full agreement with other modern models \citep{pavlov17}.
The modulation potential $\phi$ is defined here \citep[see the
full formalism in][]{usoskin_Phi_05} for the local
 interstellar spectrum according to \citet{burger00}.
Since the modulation potential is a model-dependent parameter, our result cannot be directly compared to the $\phi-$values
 based on different assumptions \citep[e.g.,][]{steinhilber12} without a recalibration \citep{herbst10,asvestari_JGR_17}.
Thus, the $\phi-$series is an intermediate result that is further converted into an physical index of the open
 magnetic flux and subsequently into the SNs.

The relation between the isotope production rate and its measured content ($\Delta^{14}$C for $^{14}$C and depositional
 flux or concentration for $^{10}$Be) depends on the corresponding atmospheric or terrestrial cycle of the isotope.
Radiocarbon is involved, as carbon-dioxide gas, in the global carbon cycle and is almost completely mixed and homogenized
 over the global hemisphere.
Here we used the globally averaged $^{14}$C production rate as computed by \citet{roth13} from the INTCAL09 standard
 $\Delta^{14}$C dataset \citep{reimer09} using a new-generation dynamic carbon cycle model that includes coupling with
 the diffusive ocean.
However, the effect of extensive fossil fuel burning (Suess effect) makes it difficult to use $^{14}$C data
 after the mid-ninteenth century because of large and poorly constrained uncertainties \citep{roth13}.
Radiocarbon data cannot be used after the 1950s because of man-made nuclear explosions that led to massive
 production of $^{14}$C.
Accordingly, we did not extend our analysis to the twentieth century \citep[cf., e.g.,][]{knudsen_GRL_09}.

In contrast to $^{14}$C, $^{10}$Be is not globally mixed, and its transport or deposition in the atmosphere is quite complicated
 and subject to local and regional conditions.
Here we applied the $^{10}$Be production model by \citet{poluianov16},
and atmospheric transport
 and deposition were considered via the parametrization by \citet{heikkila09,heikkila13}, who performed
 a full 3D simulation of the beryllium transport and deposition in the Earth's atmosphere.
However, the existing models consider only the large-scale atmospheric transport and do not address in
 full detail the deposition at each specific location; this may differ significantly from site to site.
This remains an unknown factor (up to 1.5 in either direction) between the modeled and actually measured deposition
 flux of $^{10}$Be at any given location \citep[e.g.,][]{sukhodolov17}.
On the other hand, a free conversion factor exists if the $^{10}$Be data are provided in concentration units rather
 than depositional flux.
Therefore, we considered a single scaling factor that enters the production rate $Q$ so that it matches the measured values.
This factor was
adjusted for each $^{10}$Be series separately.

Sporadic solar energetic particle (SEP) events are sometimes
produced by the Sun, with strong fluxes of energetic particles
 impinging on the Earth's atmosphere.
Although these solar particle storms usually have a short duration and soft energy spectrum, the SEP flux can produce
 additional cosmogenic isotopes in the atmosphere.
For extreme events, the enhancement of the isotope production may greatly exceed the annual yield from GCR \citep{usoskin_ApJ_12}.
If not properly accounted for, these events may mimic periods of reduced solar activity \citep{mccracken15}, since the enhanced
 isotope production is erroneously interpreted in terms of the enhanced GCR flux, and consequently, in terms of reduced solar activity.
Two extreme SEP events are known that can lead to such an erroneous interpretation \citep{bazilevskaya14}: the strongest event occurred around 775 AD \citep{miyake12}, and a weaker event was reported in 994 AD \citep{miyake13}.
The energy spectra of these events have been assessed elsewhere \citep{usoskin_775_13,mekhaldi15}.
The production effect of the event on the $^{10}$Be data in polar ice was calculated by \citet{sukhodolov17}
 and removed from the original data.
One potential candidate around 5480 BC studied by \citet{miyake17} appears to be an unusual
 solar minimum rather than an SEP event.
Accordingly, we kept it as a wiggle and did not correct for the possible SEP effect.

Here we first calculated the modulation potential $\phi$ in the past for each series individually, and then for all series together.
The reconstruction process is described in detail below.

\subsection{Reducing the series to the reference geomagnetic conditions}
\label{sec:geom}
\begin{figure}
\includegraphics[width=1\columnwidth]{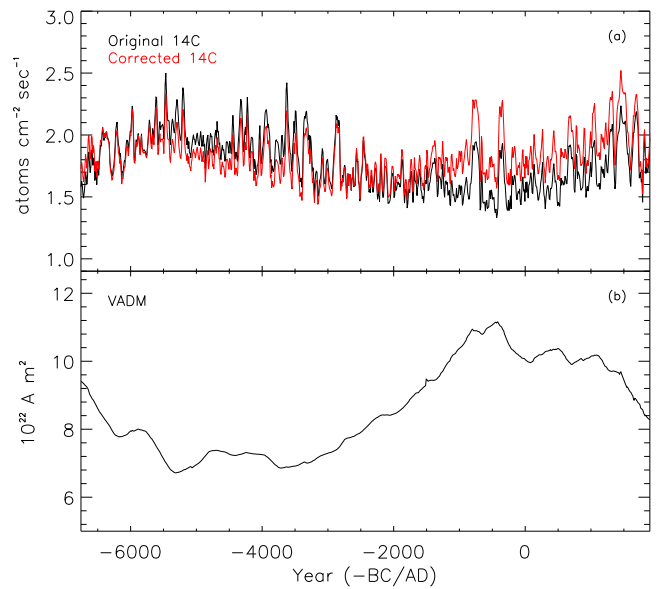}
\caption{(a) Original (black) mean $^{14}$C production rate \citep{roth13} and the one reduced to the standard
 geomagnetic conditions (red).
(b) mean VADM reconstruction \citep{usoskin_AA_16}.}
\label{Fig:geom}
\end{figure}

The temporal variability of cosmogenic isotope production contains two signals: solar modulation, and
 changes in geomagnetic field. These signals are independent of each other and can thus be separated.
Since here we are interested in the solar variability, we removed the geomagnetic signal by reducing all
 production rates to the {reference geomagnetic conditions, defined as follows: the geomagnetic field is dipole-like,
 aligned with the geographical axis, and its virtual axial dipole moment (VADM) is $8\times 10^{22}$ A m$^2$.}
The exact VADM value is not important for the procedure, therefore we chose a rounded value close to the mean value
 for the twentieth century.
The reduction was made in two steps:
\begin{enumerate}
\item[(1)]
From the given isotope production rate $Q(t)$ and the geomagnetic VADM $M(t),$ we calculated the value of
 $\phi(t)$ using the production model by \citet{poluianov16}.
\item[(2)]
From the value of $\phi(t)$ computed in step 1, we calculated the reduced production rate $Q^*(t)$
 using the same production model, but now fixing the VADM at $M=8\times 10^{22}$ A m$^2$.
This value roughly corresponds to conditions in the twentieth century.
This yields the isotope production rate as it would have been during time $t$ if the geomagnetic field had been
 kept constant at this value of $M$.
\end{enumerate}
The new $Q^*$ series is now free (in the framework of the adopted model) of the geomagnetic changes and
 is used in the subsequent reconstructions.
An example of the original series and the series corrected for variations in the geomagnetic field is shown in Figure~\ref{Fig:geom}.

\subsection{Reconstruction based solely on $^{14}$C}
\label{sec:phi_recon_14c}

First, we calculated the solar modulation potential based solely on the radiocarbon data using a Monte Carlo
 method similar to the method developed by \citet{usoskin_AAL_14,usoskin_AA_16}.
The reconstruction includes the following steps:

\begin{enumerate}

\item[(1)]
For each moment $t$ in time, we used 1000 realizations $Q_i(t)$ of the full ensemble provided by \citet{roth13}.
 These realizations include uncertainties of the carbon cycle and of the measurement errors.
At the same time, we also used 1000 realizations of the VADM $M_j(t)$ from \citet{usoskin_AA_16},
 which include uncertainties
 and cover the range of available archeomagnetic reconstructions to calculate 10$^{6}$ values of $Q_{ij}^*(t)$,
 as described in Section~\ref{sec:geom}.
The whole ensemble provides a natural way to represent the range and uncertainties
 of the reconstructed quantities ($Q$ or VADM).
For this $Q_{ij}^*(t)$ ensemble, we calculated the mean $\langle Q^*(t)\rangle$ and the standard deviation $\sigma_Q(t)$.

\item[(2)]
Using the statistics of the $Q^*(t)$ ensemble, we defined for each time $t$, the best-fit
 $\phi(t)$ that minimizes the value of $\chi^2$:
\begin{equation}
\chi^2(\phi)=\left( \frac{\langle Q^*\rangle - Q'(\phi)}{\sigma_Q} \right)^2,
\label{eq:chi2}
\end{equation}
where $Q'(\phi)$ is the value of $Q^*$ computed for a given value of $\phi$, which was scanned over the range 0 -- 2000 MeV.
The best-fit value of $\phi_0$ is defined as the value corresponding to the minimum $\chi^2_0$ ($\rightarrow 0$ for a single series used).
The 68\% confidence interval of $\phi$ is defined as the interval bounded by the values of $\chi^2=\chi^2_0+1$.
An example of the $\chi^2(\phi)$ dependence and definition of the best-fit $\phi $values and its
 uncertainties is shown in Figure~\ref{fig:14C_chi2_805}.

\begin{figure}
\centering
\includegraphics[width=1\columnwidth]{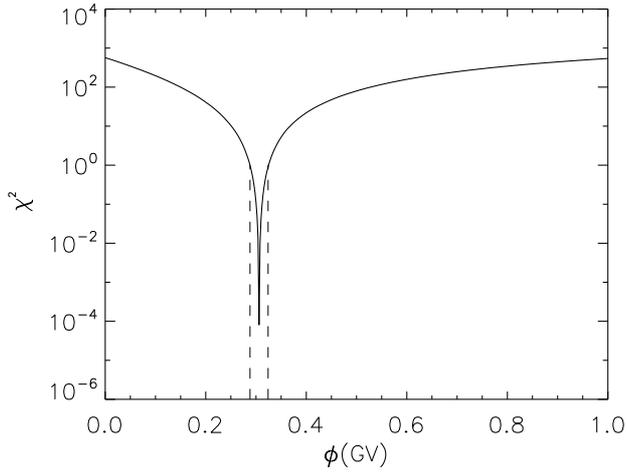}
\caption{Example of the $\chi^2$ vs. $\phi$ dependence for 805 AD for the $^{14}$C series.
The dashed lines represent the 68\% confidence interval for $\phi$.}
\label{fig:14C_chi2_805}
\end{figure}

\item[(3)]
The series of reconstructed $\phi$ based on $^{14}$C ($\phi_{\rm 14C}$) was then computed, along with the uncertainties,
 as shown in Fig.~\ref{fig:Phi_14C}.
 \end{enumerate}

\begin{figure}
\centering
\includegraphics[width=1\columnwidth]{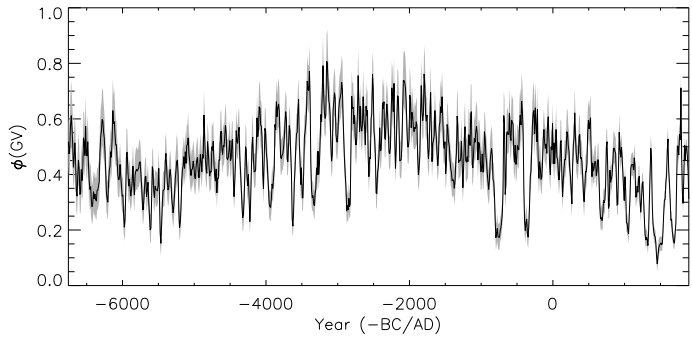}
\caption{Series of the modulation potential $\phi$ computed based only on the $^{14}$C data.
Shading denotes the 68\% confidence interval.}
\label{fig:Phi_14C}
\end{figure}

\subsection{Comparison between the long $^{10}$Be and $^{14}$C series}
\label{sec:be10_vs_C14}
\label{sec:phi_recon_grip_edml}

\begin{figure}
\centering
\includegraphics[width=1\columnwidth]{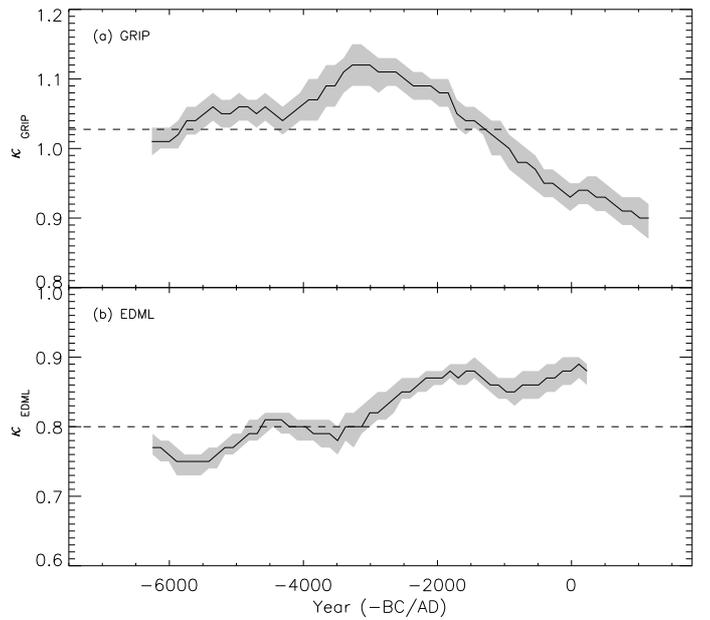}
\caption{The $^{10}$Be scaling factor $\kappa$, with 68\% uncertainties indicated by the grey shading, in the sliding 1000-year window as a function of time.
Panels a and b are for the GRIP and EDML series, respectively.
Dashed lines depict the $\kappa$ factor defined for the entire period (Section~\ref{sec:phi_recon_14C_grip_edml}).}
\label{Fig:Be_vs_C}
\end{figure}

Next we compared the long-term behavior, in the sense of the scaling factors, of the long-running $^{10}$Be series
 versus the reference $^{14}$C series.
First we considered a 1000-year window and calculated the mean value of $\langle\phi_{\rm {14}C}\rangle$ within this
 period, as described in Section~\ref{sec:phi_recon_14c}.
We then scaled each $^{10}$Be series individually with a scaling factor $\kappa$ and reconstructed
 the value of $\phi_{\kappa}$ for the rescaled $^{10}$Be series in this time window.
The value of $\kappa$ was defined such that the mean value of $\phi_{\rm 10Be}$ agrees with the value for $^{14}$C for the same period,
 that is, $\langle\phi_{\rm 10Be}\rangle=\langle\phi_{\rm14C}\rangle$.
Then, the 1000-year window was moved by 100 years and the procedure was repeated.
The resulting $^{10}$Be scaling factors $\kappa$ are shown as a function of time in Figure~\ref{Fig:Be_vs_C}.

The $\kappa$-factor for the GRIP series is relatively stable during the period 6760 BC -- 3000 BC and depicts a steady
 monotonous decrease around 3000 BC and reaching -12\% with respect to the final $\kappa_{GRIP}$ around 1000 AD.
The Pearson squared correlation between the two curves is significant, $R^2=$0.78 ($p-$value 0.02)
 \footnote{The significance is estimated using the non-parametric random-phase method by \citet{ebisuzaki97}.} , implying
 a possible residual effect of the geomagnetic field in the reconstruction.

It is interesting to note that the two $^{10}$Be series show very different trends during the second half of the Holocene.
The $\kappa$-factor for the EDML series shows a weak growing trend against the $^{14}$C series over the entire
 period of their overlap, with $\kappa$ varying from -5\% to +10\%.
A shallow wavy variability can be noticed, with a quasi-period of approximately 2400 years, which is
 probably related to the Hallstatt cycle \citep{damon91,usoskin_AA_16}.
The correlation between the $\kappa-$factor for EDML and the VADM is insignificant $R^2=0.52$ ($p=0.12$).
This suggests that the under-corrected geomagnetic field effect is most likely only related to the GRIP series.

The discrepancy between GRIP and $^{14}$C series is well known \citep[e.g.,][]{vonmoos06, inceoglu15},
 but has typically been ascribed to the early part of the Holocene because both series are normalized to the modern period.
With this normalization, the records agree with each other over the last millennia but diverge before
 ca. 2000 BC \citep{inceoglu15}.
This discrepancies has sometimes \citep[e.g.,][and references therein]{usoskin_LR_17} also been explained as a
 possible delayed effect (not perfectly stable thermohaline circulation) of the deglaciation in the carbon cycle
 \citep[e.g.,][]{muscheler04}.
However, as we show here, this explanation is unlikely for two reasons.

\begin{figure}[th]
\includegraphics[width=0.9\columnwidth]{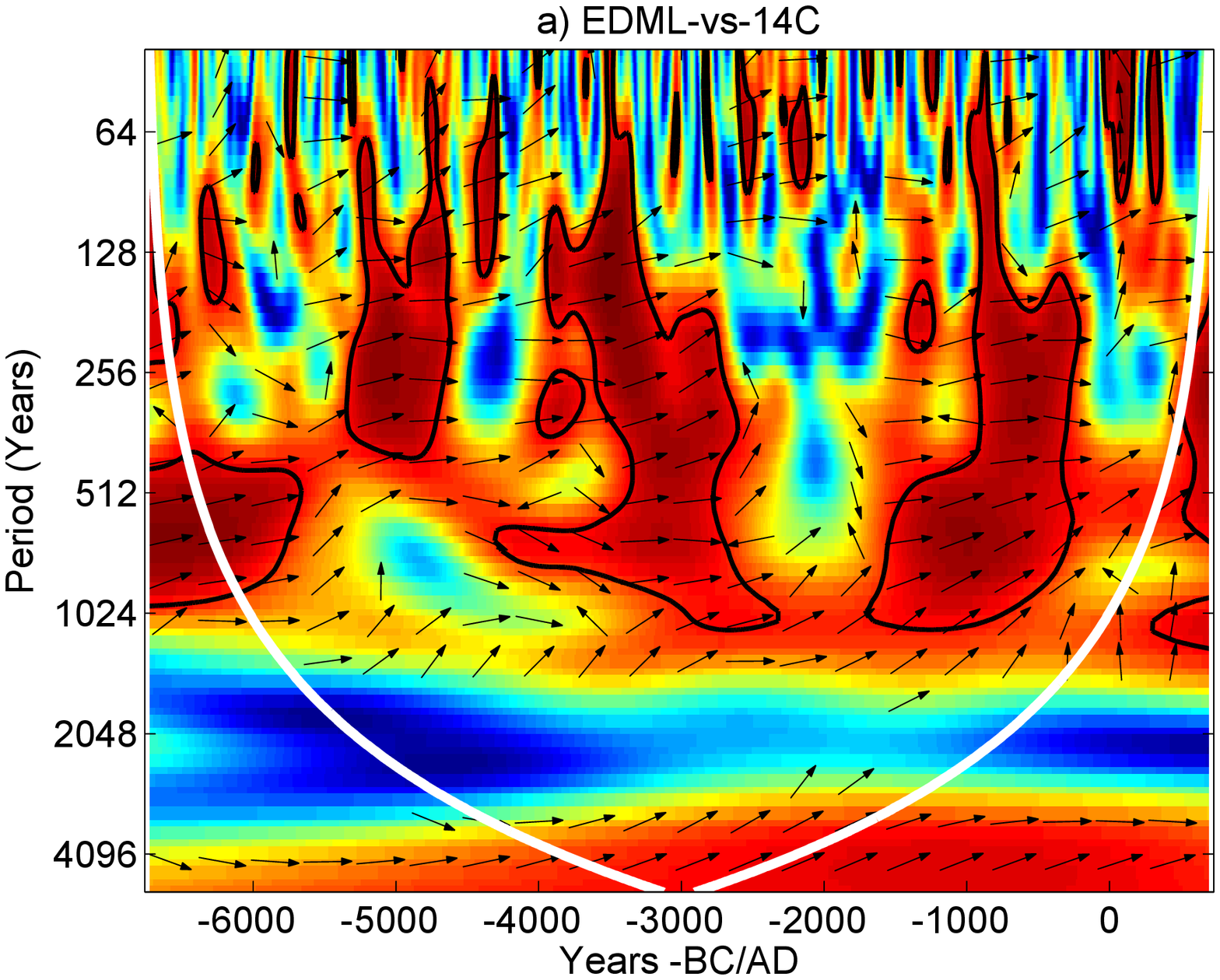}
\vskip 0.1cm
\includegraphics[width=0.9\columnwidth]{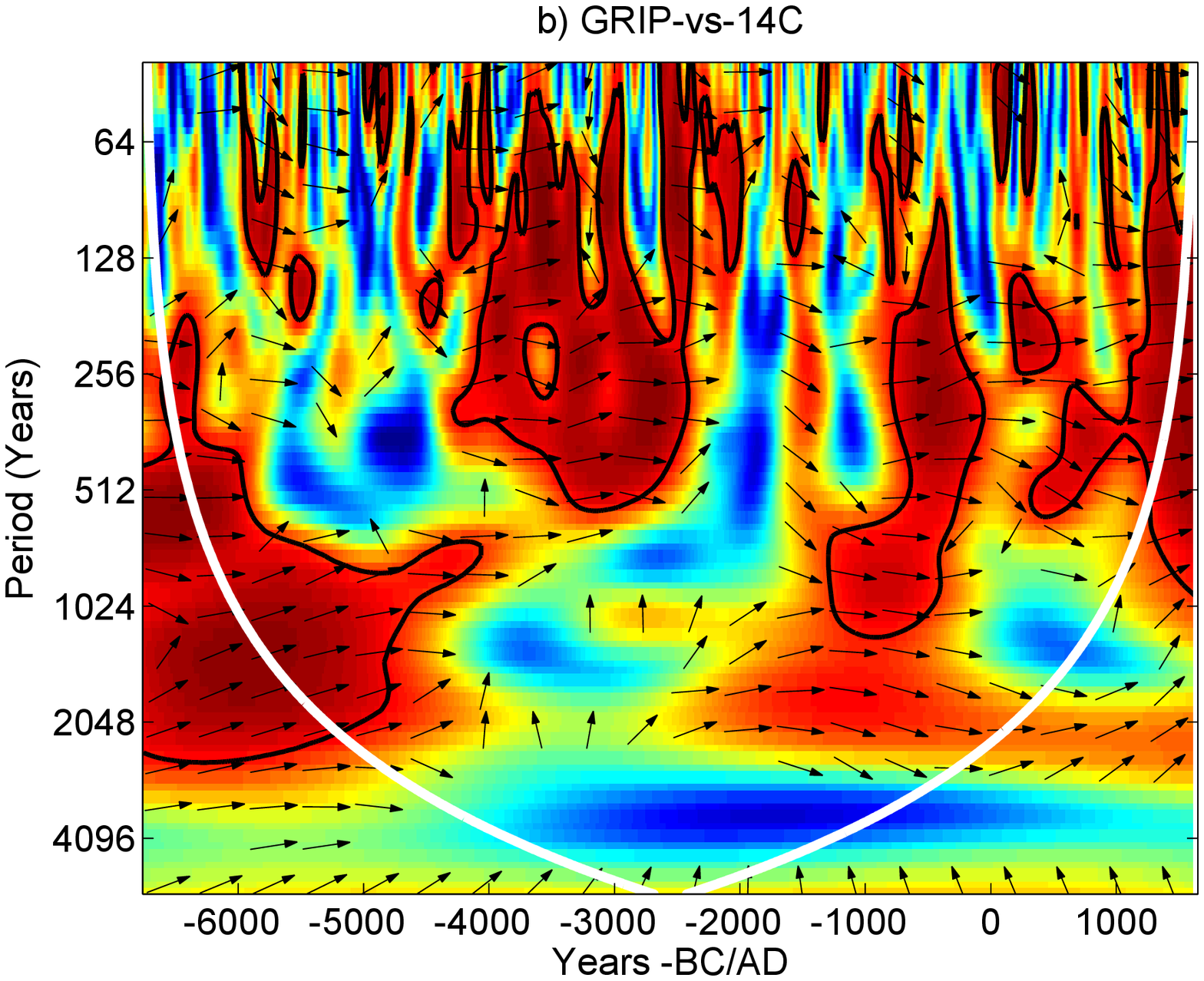}
\vskip 0.1cm
\includegraphics[width=0.9\columnwidth]{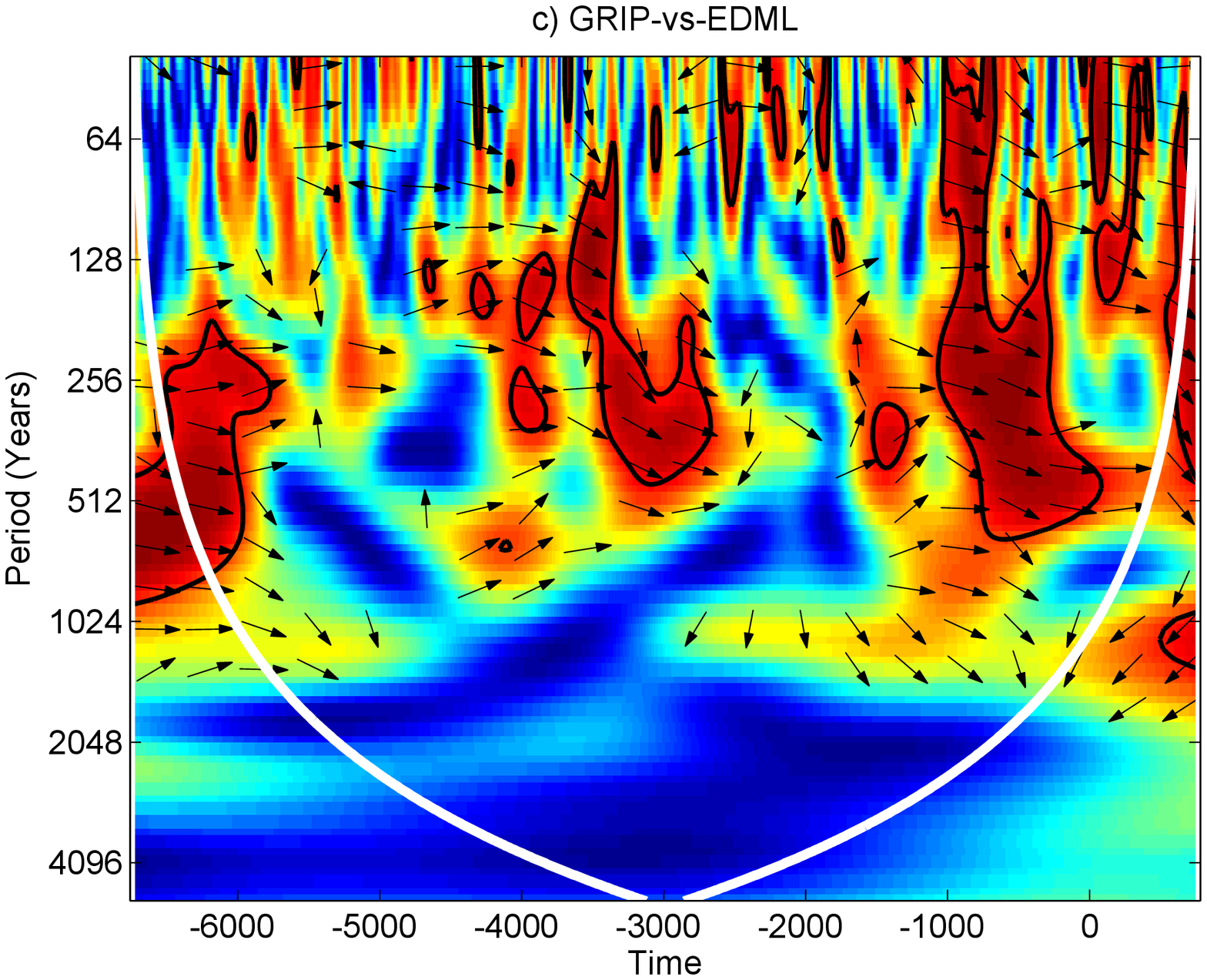}
\includegraphics[width=0.09\columnwidth]{scale.eps}
\caption{Wavelet coherence between individual long-term series of the modulation potential $\phi$
 reconstructed from individual long-term cosmogenic series (shown in Figure~\ref{fig:Phi_k}).
(a) $^{10}$Be EDML vs. $^{14}$C;
(b) $^{10}$Be GRIP vs. $^{14}$C;
(c) $^{10}$Be GRIP vs. $^{10}$Be EDML.
Notations are similar to Figure~\ref{fig:ww_a}.}
\label{fig:ww}
\end{figure}

More details of the long-term relation between the series are shown in Figure~\ref{fig:ww}, where we plot
 the pairwise wavelet coherence between the modulation potential (shown in Figure~\ref{fig:Phi_k}) from the
 three long series.
Panel a shows that the EDML series is coherent with the $^{14}$C series at all timescales shorter than
 $\approx 2000$ years and again (but insignificantly) longer than 4000 years, while no coherence exists
 between 2000 and 4000 years.
The GRIP-based series (panel b) is well coherent with the $^{14}$C one on timescales longer than 100--200
 year, and the coherence disappears at timescales longer than 2000--3000 years.
The two $^{10}$Be are hardly coherent with each other on timescales longer than 1000 years (panel c),
 with a small insignificant isle of coherence beyond the cone of influence.
It is noteworthy that each of the $^{10}$Be series exhibits a higher coherence with the $^{14}$C one than with each other.
A similar conclusion was made by \citet{usoskin_10Be_09} using different $^{10}$Be data series.
A reason for this discrepancy is yet to be discovered.

Since the reason for this discrepancy is unknown, we did not correct for it.
In the following we apply fixed coefficients for $^{10}$Be series, thus keeping
 the long-term variability as it exists in the original data.

It is interesting that while the coherence between EDML- and $^{14}$C-based reconstructions
 has been improved compared to the coherence between the raw data series (Figure~\ref{fig:ww_a}),
 the coherence is degraded for the GRIP series (panels b) in timescales longer than
 2000 years.
This suggests that beryllium may be better mixed for Greenland than was derived through the applied beryllium
 transport parametrization based on \citet{heikkila09}.
The reconstruction based on the $^{14}$C series has no residual correlation with VADM ($R^2<0.01$).
A new full-size model of the atmospheric transport and deposition of $^{10}$Be in polar regions,
 particularly in Greenland, is needed  to resolve the issue.

\subsection{Reconstructions from individual series}

Two beryllium series, GRIP and EDML, were used here for a preliminary assessment of this method for their overlap.
The mean modulation potential for the period 6800 BC -- 700 AD based on radiocarbon is
 $\langle\phi_{\rm 14C}\rangle=0.468$ GV.

First, we scaled each $^{10}$Be series individually with a free scaling factor $\kappa_0$ and reconstructed the
 series of $\phi$.
The values of $\kappa_0$ were defined such that the mean value $\langle\phi_{\rm 10Be}\rangle$ is equal
 to that for $^{14}$C for the same period.
The scaling factor we found for the GRIP data is close to unity, $\kappa_0=1.1$.
This implies that $^{14}$C and $^{10}$Be data are fully consistent in the framework of the model, and that our
 modeling of the cosmogenic isotope production and transport
or deposition is appropriate.
The best-fit scaling factor for EDML is $\kappa_0=0.8$, which
is about 20\% lower than unity, which is reasonable considering
 specific features of the deposition at this site.
These values of $\kappa_0$ were considered as initial guesses for a more precise search for the scaling factors,
 as described below.

%

\subsection{Combined-record reconstruction of $\phi$}
\label{sec:phi_recon_14C_grip_edml}

In the preceding section we calculated three series of $\phi$ from different cosmogenic isotope datasets,
 using their inter-calibration to the mean $\phi$ value obtained from the $^{14}$C data over the overlap period.
This is similar to what has been done previously \citep{vonmoos06,steinhilber12,usoskin_AAL_14,usoskin_AA_16}.
Here we proceed and perform a consistent {Bayesian-based reconstruction of the
 modulation potential $\phi$.
Using the measured data and knowledge of the other complementary parameters,
 we determine for each moment in time the most probable value of $\phi$ and its uncertainty.}

\begin{enumerate}
\item[(1)]
First, we fixed the scaling factors for the GRIP and EDML series at their initial guess values $\kappa_0$ as described above.

\item[(2)]
We then calculated, as described in Section~\ref{sec:phi_recon_14c} (step 1), $10^6$ realizations of the isotope
 production rates $Q^*(t)$ reduced to the standard geomagnetic conditions.

\item[(3)]
The mean $\langle Q^*(t)\rangle$ and the standard deviation $\sigma_{Q^*}(t)$
 were calculated over the $10^6$ ensemble members for each time point $t$.

\item[(4)]
For each $t,$ the value of $\chi^2(\phi)$ was calculated as
\begin{equation}
\chi^2(\phi)=\sum_{i=1}^3\left( \frac{\langle Q_i^*\rangle - Q_i'(\phi)}{\sigma_{Q_i}} \right)^2,
\label{eq:chi2_n}
\end{equation}
where the index $i$ takes values 1 through 3 for the $^{14}$C, GRIP, and EDML series, respectively.
An example of $\chi^2$ as a function of $\phi$ at one point in time for the three individual series as well
 as for their sum is shown in Figure~\ref{fig:chi2_-2095}.
Each of the individual datasets (panels a--c, each similar to Figure~\ref{fig:14C_chi2_805}) yields a very
 sharp and well-defined dip in the $\chi^2$ value; this dip approaches zero.
The reason is that for a given single value of $Q$, the corresponding value of $\phi$ can be defined precisely.
However, the obtained individual $\phi-$values are not identical for different datasets, which leads to a smooth overall
 $\chi^2$-vs-$\phi$ dependence (panel d), the minimum $\chi^2$ of which is about 1.7 or 0.85 per degree of freedom (DoF).
This implies that the same value of $\phi=0.58$ GV satisfies all three isotope data records within statistical confidence.
\begin{figure}
\centering
\includegraphics[width=1\columnwidth]{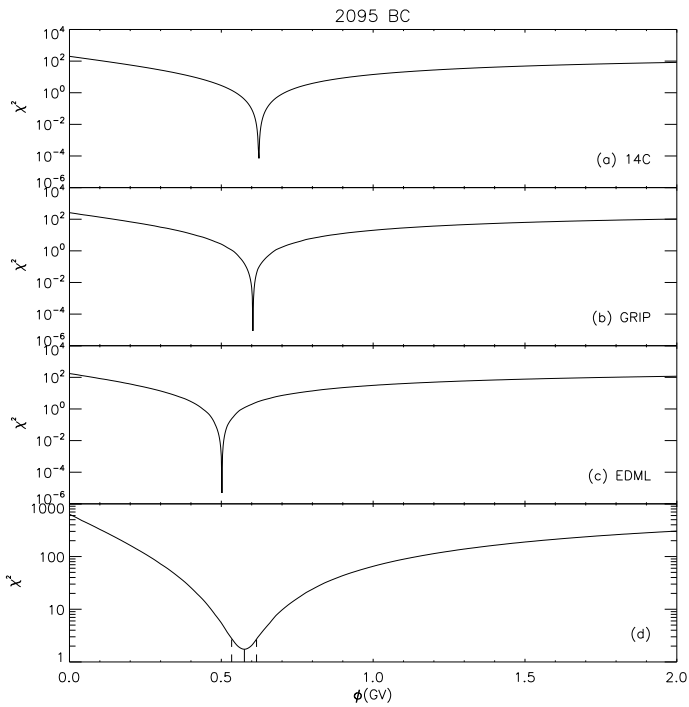}
\caption{Dependence of $\chi^2$ (Eq.~\ref{eq:chi2_n}) on $\phi$ for the decade centered at 2095 BC: (a) $^{14}$C
 with no scaling,
(b) $^{10}$Be GRIP with scaling ($\kappa=1.028$), (c) EDML with scaling $\kappa=0.815$, and (d) sum of the three
 $\chi^2$ components.
}
\label{fig:chi2_-2095}
\end{figure}

\item[(5)]
We calculated the sum of individual $\chi^2$ values (Eq.~\ref{eq:chi2_n}) as $\chi_\Sigma^2=\sum_t\chi^2(t)$ over
all 750 time points during 6760 BC -- 730 AD.
The corresponding sum is $\chi^2_\Sigma=4347$ or $\approx 2.9$ per DoF, indicating a likely systematic difference between the series.
The DoF number is defined as $750\times 3$ (number of points in the three series) minus 752 (the number of
 fitted parameters), thus 1498.

\item[(6)]
Since the values of the scaling factors $\kappa_0$ were initially defined by normalizing the mean values of $\phi$,
 which is not optimal, we redefined them via $\chi^2$ as follows.
We repeated steps (1)--(5) above, now scanning the values of $\kappa$ in the range of 0.85--1.25 and 0.6--1.1, with steps 0.01 and 0.015,
 for the GRIP and EDML series, respectively.
The corresponding $\chi^2_\Sigma$ were calculated.
The distribution of $\chi^2_\Sigma$ as a function of $\kappa_{\rm GRIP}$ and $\kappa_{\rm EDML}$ is shown
 in Figure~\ref{fig:chi2D}.
The distribution has a clear minimum ($\chi^2_{\rm min}=3205$), and the values of $\kappa_{\rm GRIP}$ and $\kappa_{\rm EDML}$
 are 1.028 and 0.815, respectively, which is close to the initial guess (Section~\ref{sec:phi_recon_grip_edml}).
 \end{enumerate}
\begin{figure}
\centering
\includegraphics[width=1\columnwidth]{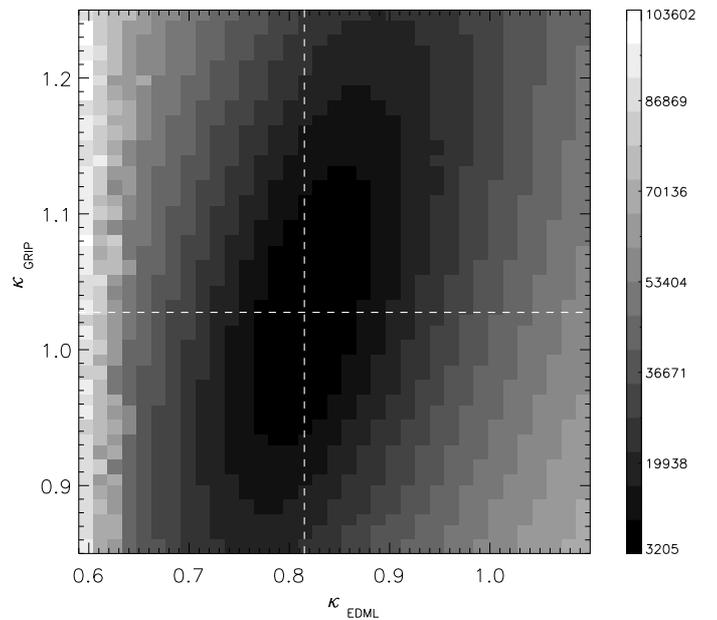}
\caption{Distribution of $\chi_\Sigma^2$ versus  $\kappa_{\rm GRIP}$ and $\kappa_{\rm EDML}$.
The gray scale is on the right.
The minimum of the distribution ($\chi^2_{\rm min}=3205$) occurs at $\kappa_{\rm GRIP}$=1.028 and $\kappa_{\rm EDML}$=0.815,
 as marked by the cross-hairs.}
\label{fig:chi2D}
\end{figure}

\begin{figure}
\centering
\includegraphics[width=1\columnwidth]{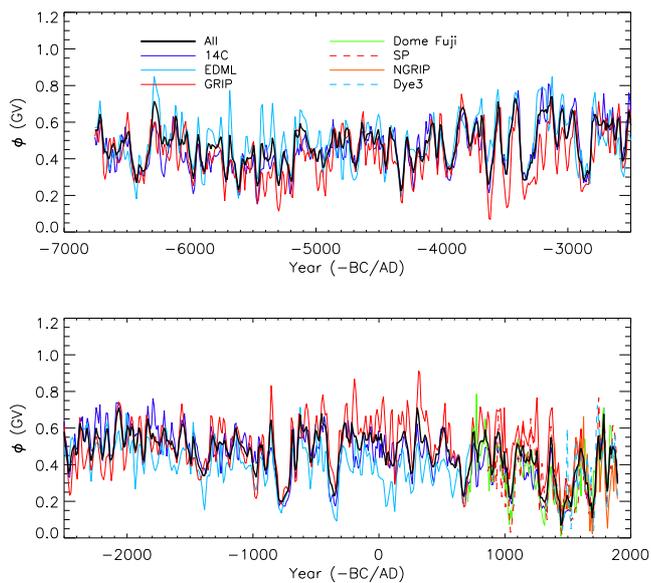}
\caption{ Reconstruction of the modulation potential $\phi$ using only individual cosmogenic isotope series
 (color curves as denoted in the legend)
 with the best-fit scaling (see Table~\ref{tbl:series}) and the final composite series (thick black curve).
 The top and bottom panels depict the two halves of the entire interval.
 Only mean values are shown without uncertainties.}
\label{fig:Phi_k}
\end{figure}

Reconstructions of $\phi$ based on the best-fit scaling for individual series are shown in Figure~\ref{fig:Phi_k} as colored curves.
{The differences between the data points from the different series are normally distributed, with a mean value of about zero
 and the standard deviation of 100--200 MV.
This serves as an estimate of the accuracy of individual reconstructions.}

\subsection{Full reconstruction}
\label{sec:phi_recon_all}

Next, we analyzed all the available data series, also including four shorter $^{10}$Be series. This extends
 the $\phi$ reconstruction to 1900 AD.
The four shorter series are (see Table~\ref{tbl:series}) NGRIP and Dye3 from Greenland, and Dome
 Fuji (DF) and South Pole (SP) from Antarctica.

\begin{figure}
\centering
\includegraphics[width=1\columnwidth]{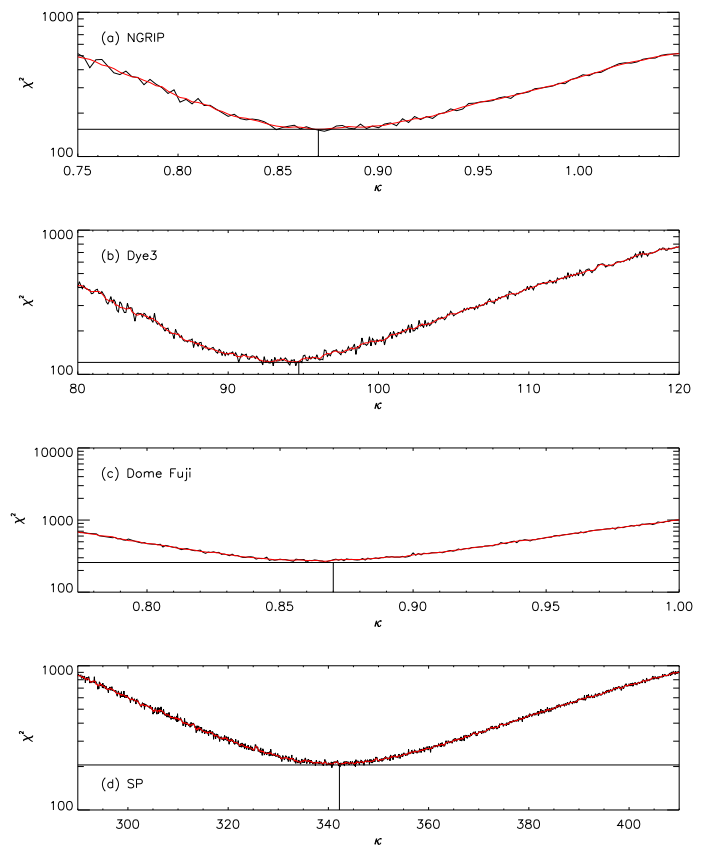}
\caption{Values of $\chi^2$ versus the scaling factors $\kappa$ for four short $^{10}$Be series, as marked in each panel.
The five-point running mean curves are shown in red.
The locations of the best-fit $\kappa-$values are shown as straight black lines.}
\label{fig:scalfacs}
\end{figure}

We first estimated the best-guess scaling $\kappa-$factors for each short series similar to what is described in
 Section~\ref{sec:phi_recon_grip_edml}, that is, we equalized the mean values of $\langle\phi\rangle$ for the series in
 question to the value of the reference $^{14}$C series for the period of their overlap.
We then slightly varied the $\kappa-$value around these best-guessed values to calculate the corresponding $\chi^2$ for
 each individual series, as shown in Figure~\ref{fig:scalfacs}.
The vertical solid lines mark the best-fit $\kappa$-values that minimize the five-point smoothed $\chi^2$.
The best-fit scaling factors are listed in Table~\ref{tbl:series}.
The scaling factor for $^{10}$Be series with depositional flux data (GRIP, EDML, NGRIP, and Dome Fuji,
 see Table~\ref{tbl:series} and Figure~\ref{Fig:Be_vs_C}) are
 close to unity (within 20\%), which again implies that our model is quite realistic.
The 68\% uncertainties are defined as corresponding to $\left(\chi^2_{\rm min}+1\right)$.

Next, we performed the full reconstruction using the $\chi^2$ method described in Section~\ref{sec:phi_recon_14C_grip_edml},
 but now for all the series (see Fig.~\ref{fig:Phi_k}).
The number of different series used to reconstruct individual data points varied in time between two and five
 (see Figure~\ref{fig:series}).
The mean $\chi^2$ per DoF is 0.57, for 84\% of data points $\chi^2<1$ per DoF, implying that
 the agreement between different series is good.

\section{Reconstruction of the sunspot number}
\label{sec:SN}

\begin{figure*}
\centering
\includegraphics[width=1\textwidth]{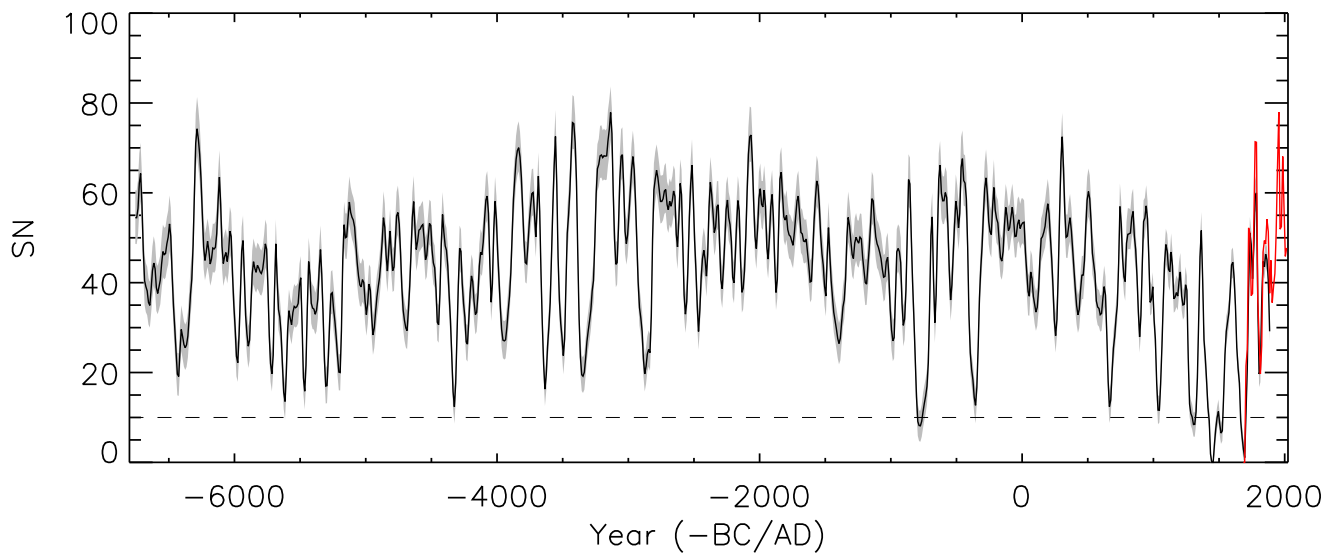}
\caption{Reconstructed sunspot number along with its 68\% confidence interval (gray shading).
This series is available in the ancillary data.
The red line depicts the decadally resampled international sunspot number (version 2, scaled by 0.6) from \citet{clette14}.
The dashed line denotes the level of SN=10.}
\label{fig:SN}
\end{figure*}

The modulation potential series alone are not very useful as a solar activity proxy since the modulation potential
 is a relative index whose absolute value is model dependent \citep{usoskin_Phi_05,herbst10,herbst17}.
Therefore, we converted the modulation potential, reconstructed in Section~\ref{sec:phi_recon}, into a more definitive index,
 the SN$\text{}\text{}$.
This was done via the open solar magnetic flux $F_{\rm o}$, following an established procedure
 \citep[e.g.,][]{usoskin_PRL_03,solanki_Nat_04,usoskin_AA_16}.
Applying the updated SATIRE-M model \citep[]{vieira10,vieira11,wu18}, we can write the relation
 \citep[][see the Appendix therein]{usoskin_AA_07} between the two indices as
\begin{equation}
{\rm SN}_{i}= 116\times \phi_{i} + 33\times(\phi_{i+1} - \phi_{i}) - 16,
\label{eq:fo2ssn}
\end{equation}
where the SN during the $i$th decade, SN$_i$, is defined by the modulation potential
 (expressed in GV) during the contemporary and the following decades.
The negative offset term (-16) reflects the fact that zero SN during grand minima does
 not imply the absence of GCR modulation, so that even when $SN=0,$ the value of $\phi$ is about 0.14 GV \citep[e.g.,][]{owens12}.
Since SNs cannot be negative, the SN was assigned zero values when the values of $\phi$ dropped below the
 sunspot formation threshold.
The uncertainties of the reconstructed $SN  $ values were defined by converting the low- and upper-bound (68\%)
 $\phi \ $values (defined as described above) for each decade into SNs.
The reconstructed SN series
 \footnote{Available as a table in the supplementary materials and at the MPS sun-climate web-page
  http://www.mps.mpg.de/projects/sun-climate/data}
 is shown in Figure~\ref{fig:SN} along with its 68\% confidence interval and is compared with the international SN series \citep[SILSO ISN version 2,][]{clette14}\footnote{File SN\_y\_tot\_V2.0.txt available at http://www.sidc.be/silso/ infosnytot}.
Since we used SNs in their `classical' definition, the ISN v.2 data were scaled down by a
 factor 0.6 \citep[as described in][]{clette14}.

The reconstructed solar activity varies at different timescales from decades to millennia.
In particular, a long period with relatively high activity occurred
between roughly 4000 BC and 1500 BC,
 and periods of lower activity occurred at ca. 5500 BC and 1500 AD.
The origin of this is unclear.
For example, \citet{usoskin_AA_16} suggested, based on the fact that the GRIP and $^{14}$C series behave differently at
 this timescale, that it is a global climate effect rather than solar.
As a result, they removed this wave from the data in an ad hoc manner.
However, as we show here, it is more likely related to the undercorrection of the GRIP-based series and
therefore may be related to solar activity, so that we retained it in the final dataset.

The new series is generally consistent with previous reconstructions \citep[e.g.,][]{usoskin_AA_16}, but
 it also has some new features.
In particular, it implies a lower activity during the sixth millennium BC.
We note that a possible overestimation of the  solar activity for this period has been suspected
 previously \citep{usoskin_AA_07,usoskin_AA_16}.

Although the overall level of solar activity may vary significantly, the periods of grand minima may
 correspond to a special state of the solar dynamo \citep{schmitt96,kueker99,moss08,choudhuri12,kapyla16}
 and thus are expected to provide roughly
 the same low level of activity, corresponding to a virtual absence of sunspots.
Thus, the level of the reconstructed activity during clearly distinguishable grand minima may serve as a rough
 estimate of the `stability' of the reconstruction when we assume that activity always drops
 to the same nearly zero level during each grand minimum \citep{sokoloff94,usoskin_AAL_14}.
The expected level of solar activity during grand minima (SN$\rightarrow$0) is indicated by the horizontal dashed
 line in Figure~\ref{fig:SN}.
The level of the reconstructed grand minima (observed as sharp dips) is roughly consistent with
 SN$\rightarrow$0 throughout almost the entire period, considering an uncertainty on the order of 10 in SN units.
However, periods of 3500 BC -- 2500 BC and before 5500 BC are characterized by a slightly higher SN level during
 the grand minima, suggesting a possible overestimation of activity during these times.
Interestingly, no clear grand minima occurred during 2500 BC -- 1000 BC, suggesting that it was a long period of
 stable operation of the solar activity main mode.

The overall level of solar activity remained roughly constant, around 45--50, during most of the time,
 but appeared somewhat lower (around 40) before 5000 BC, suggesting a possible slow variability
 with a timescale of about 6--7 millennia \citep[cf.,][]{usoskin_AA_16}.
The reason for this is unknown, but it might be due to
(1) climate influence, although this is expected to affect $^{14}$C and $^{10}$Be isotopes differently;
(2) solar activity, or
(3) large systematic uncertainty in the geomagnetic field reconstruction, which is poorly known before about 3000 BC.

\begin{figure}
\centering
\includegraphics[width=\columnwidth]{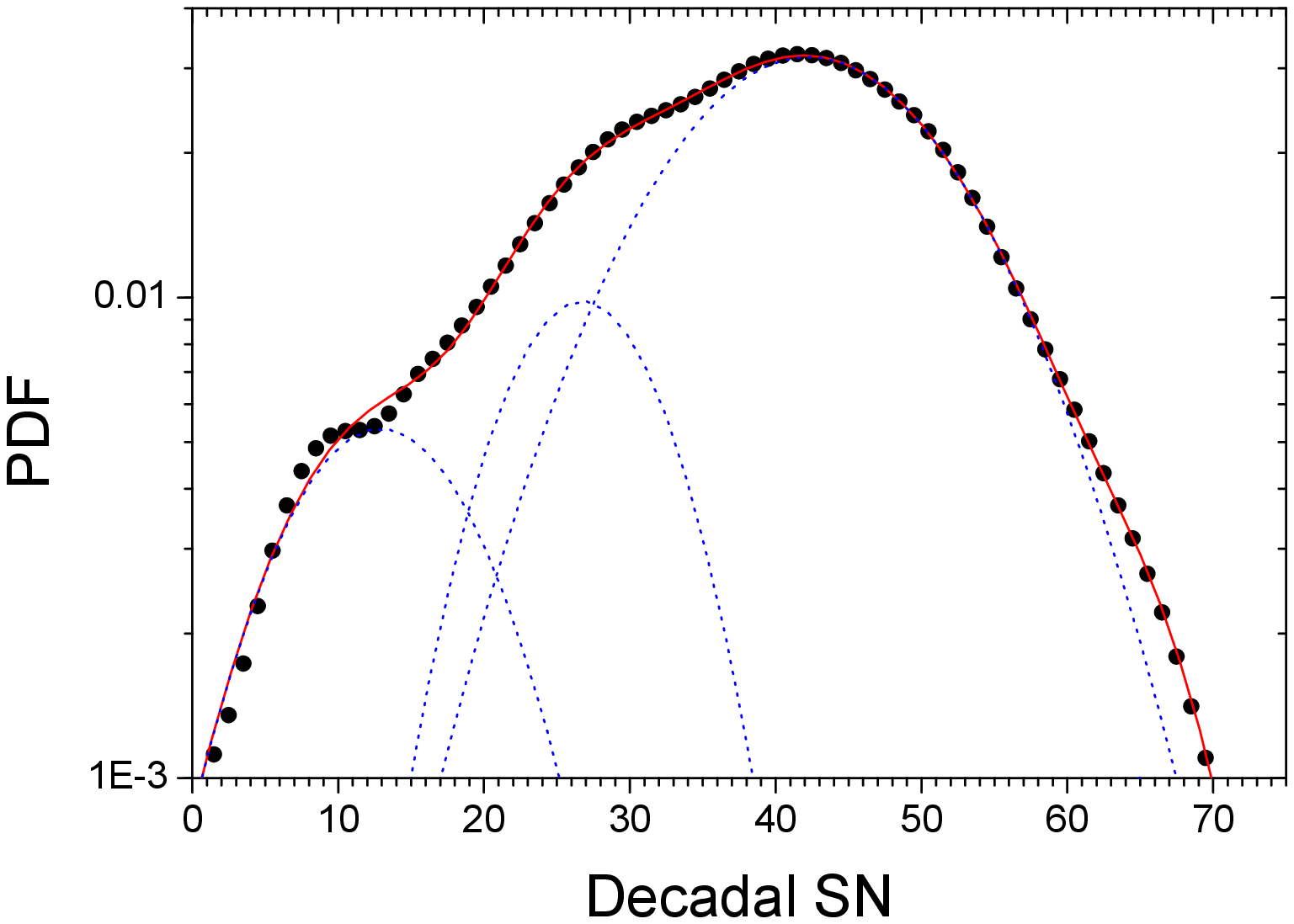}
\caption{Probability kernel density function estimate (black dots) of the reconstructed decadal sunspot numbers
 (Gaussian kernel, width=3) and its multi-peak fit (red).
}
\label{fig:kdf}
\end{figure}

Figure~\ref{fig:kdf} shows the kernel density estimation of the probability density function (KDF) of the
 reconstructed decadal SNs for the entire period (877 decades).
The high peak of the distribution with values around 40 is clearly
visible; this corresponds to moderate activity.
A low-activity component with values below 15 is clearly visible
as well; this corresponds to a grand minimum.
In addition, a bump is faintly visible at the high-activity tail (SN greater than 60, visible only
 as a small excess above the main Gaussian curve), suggesting a grand maximum component.
In order to illustrate this, we applied a formal multi-peak (Gaussian) fit to the KDF, which is shown as the
 blue dotted curves.
The main peak is centered at decadal SN=42 ($\sigma=10$) and represents the main component.
Another peak can be found as a small bump around SN=27 ($\sigma=6$), but it is not sufficiently separated from the
 main component to justify calling it a separate component.
The grand minimum component (Gaussian with $\sigma=6$ centered at SN=12) is significantly separated from the
 main component (statistical significance $p<0.05$).
The separation of the grand maximum component, while visible by eye as a deviation from the Gaussian shape at high values,
 is not statistically significant.
The statistical separation of the special grand minimum component was first shown by
 \citet{usoskin_AAL_14} for the last three millennia, while the result for the grand maximum component was inconclusive.

Here we fully confirm this result over nine millennia, which implies that grand minimum and normal activity components form a
 robust feature of solar variability.
This also suggests that the new reconstruction is more robust and less noisy than previous reconstructions
 \citep[e.g.,][]{steinhilber12,usoskin_AA_16} and allows
 statistically identifying the grand minimum component over 9000 years.

\section{Conclusions}

We have provided a new fully consistent multi-proxy reconstruction of the solar activity over about nine millennia,
 based for the first time on a Bayesian approach.
We used all the available datasets of cosmogenic radioisotopes with sufficient length and quality in terrestrial
 archives and up-to-date models of isotope production and transport
or deposition as well as a recent archeomagnetic model.
We used six $^{10}$Be series of different lengths from Greenland and Antarctica, and the official INTCAL
 global $^{14}$C series.
Earlier reconstructions were based on either individual datasets or on a statistical superposition of them
 \citep[e.g.,][]{steinhilber12}.
Our new method is based on finding the most probable value of the solar modulation potential
 that matches all the data for a given point in time, providing also a straightforward estimate of the uncertainties.
Prior to the analysis, long $^{10}$Be series were formally redated to match wiggles in the $^{14}$C data.
All employed cosmogenic  isotope series were reduced to the reference geomagnetic field conditions.

We have also tested the stability of the two long $^{10}$Be series and found that they appear to diverge
 from each other during the second half of the Holocene, while the $^{14}$C series lies in between them.
The GRIP-based series appears to anticorrelate with the long-term geomagnetic VADM series,
 implying a possible undercorrection for the geomagnetic shielding effect for the GRIP location, while
 EDML and $^{14}$C-based reconstructions do not show any significant residual correlation with VADM.
This suggests that the applied model of beryllium transport and deposition does not work properly
 for the Greenland site but is reasonably good for the Antarctic site.
A full-size transport and deposition model \citep[e.g.,][]{sukhodolov17} needs to be applied in the
 future to resolve this issue.

The reconstructed series (Figure~\ref{fig:SN}) shows variability on different timescales.
Grand minima of activity are of particular interest. They are  visible as strong dips in the time series, in which the
 level of the decadal SNs sinks below 10--15.
Another feature of the long-term evolution of the solar activity is a slow variability on the 6--7-millennia timescale with
 lows occurring in ca. 5500 BC and 1500 AD.
This behavior has been interpreted by \citet{usoskin_AA_16} as a possible effect of climate influence on the carbon cycle,
 but the $^{14}$C series lies between the two diverging $^{10}$Be series on this timescale, which
 makes this explanation unlikely.
The cause of the feature remains unknown.

Most of the time, the solar activity varies slightly around the moderate level of SN$\approx 40$, which corresponds
 to the main component of the solar activity.
Grand minima form a statistically distinguishable component,
however.
The existence of this component was first determined for the last three millennia by \citet{usoskin_AAL_14},
 who interpreted it as special mode of the solar dynamo.
Its confirmation here for nine millennia implies that it is a robust feature of the solar activity.
At the same time, the possible existence of a component representing grand maxima is indicated
 \citep[cf.][]{usoskin_AAL_14}, although
 it cannot be separated from the main component in a statistically significant manner.
We speculate that the different components of the activity distribution may be related to different modes
 of the dynamo operation.

Finally, a new consistent reconstruction of the solar activity (in the form of decadal SNs) was presented
 that offers a more reliable estimate of the long-term evolution of the solar variability and poses robust constraints on
 the development of solar and stellar dynamo models as well as solar-terrestrial studies.

\begin{acknowledgements}
We acknowledge Mads Knudsen for consultations on the GRIP series in GICC05 timescale and Raimund Muscheler for discussions
 about the ice core chronologies.
Support by the Academy of Finland to the ReSoLVE Center of Excellence (project no. 272157) is acknowledged.
C.J.W. acknowledges postgraduate fellowship of the International Max Planck Research School for Solar System Science.
This work has been partly supported by the BK21 plus program through the National Research Foundation (NRF)
 funded by the Ministry of Education of Korea.
\end{acknowledgements}


\end{document}